%% file: main.tex
\relax
\documentclass[letterpaper]{article} % DO NOT CHANGE THIS
\usepackage{aaai22}  % DO NOT CHANGE THIS
\usepackage{times}  % DO NOT CHANGE THIS
\usepackage{helvet}  % DO NOT CHANGE THIS
\usepackage{courier}  % DO NOT CHANGE THIS
\usepackage[hyphens]{url}  % DO NOT CHANGE THIS
\usepackage{graphicx} % DO NOT CHANGE THIS
\usepackage{tabularx}
\usepackage{booktabs}
\usepackage{subfig}
\usepackage{multirow}
\usepackage{amsmath}
\usepackage[T1]{fontenc}
\usepackage{float}
\usepackage{dblfloatfix} 
\usepackage{natbib}  % DO NOT CHANGE THIS AND DO NOT ADD ANY OPTIONS TO IT
\usepackage{caption} % DO NOT CHANGE THIS AND DO NOT ADD ANY OPTIONS TO IT
\DeclareCaptionStyle{ruled}{labelfont=normalfont,labelsep=colon,strut=off} % DO NOT CHANGE THIS
\usepackage{algorithm}
\usepackage{algorithmic}

\usepackage{newfloat}
\usepackage{listings}
\floatstyle{ruled}
\newfloat{listing}{tb}{lst}{}
\floatname{listing}{Listing}

\title{Extending Cross-Modal Retrieval with Interactive Learning to Improve Image Retrieval Performance in Forensics}
\author{
    Nils Böhne,\textsuperscript{\rm 1}
    Mark Berger,\textsuperscript{\rm 2}
    Ronald van Velzen\textsuperscript{\rm 2}
}
\affiliations {
    \textsuperscript{\rm 1} University of Amsterdam\\
    \textsuperscript{\rm 2} FIOD (Fiscal Information and Investigation Service)\\
    nilsbohne@outlook.com, mark@maberger.nl, ronald\_vanvelzen@hotmail.com
}

\begin{document}

\maketitle

\begin{abstract}
Nowadays, one of the critical challenges in forensics is analyzing the enormous amounts of unstructured digital evidence, such as images. Often, unstructured digital evidence contains precious information for forensic investigations. Therefore, a retrieval system that can effectively identify forensically relevant images is paramount. In this work, we explored the effectiveness of interactive learning in improving image retrieval performance in the forensic domain by proposing Excalibur - a zero-shot cross-modal image retrieval system extended with interactive learning. Excalibur was evaluated using both simulations and a user study. The simulations reveal that interactive learning is highly effective in improving retrieval performance in the forensic domain. Furthermore, user study participants could effectively leverage the power of interactive learning. Finally, they considered Excalibur effective and straightforward to use and expressed interest in using it in their daily practice.
\end{abstract}

\input{sections/introduction}
\input{sections/statusquo}

\input{sections/related-work}
\input{sections/system-overview}
\input{sections/experimental-setup}
\input{sections/results}
\input{sections/path_to_deployment}
\input{sections/discussion}
\input{sections/conclusion}

\bibliography{aaai22}

\end{document}

%% file: sections/introduction.tex
\section{Introduction} White-collar crime has become more widespread and costly than ever before and poses a substantial threat to the development and stability of economies \cite{hasham2019financial}. In 2021, the Dutch Fiscal Information and Investigation Service (FIOD) conducted over 900 investigations, yielding a total of more than 520 million euros in transactions for the public treasury\footnote{https://www.fiod.nl/jaarbericht-fiod-2021-financieel-en-fiscaal-rechercheren-in-een-digitale-internationale-wereld/}. Therefore, combating fiscal crime continues to be one of the highest priorities of governments worldwide in the field of law enforcement.

Meanwhile, the amount of digital evidence continues to increase far beyond the capacities of investigators to analyze and process effectively \cite{battiato2012multimedia}. With most mobile devices equipped with cameras, images often comprise a large portion of the digital evidence. These images may contain valuable information for the investigation. Still, most information systems for digital forensics are limited to presenting images as a sequential list of files that have to be manually inspected \cite{worring2012multimedia}. Naturally, this negatively impacts the timely analysis of digital evidence in an investigation while also affecting the timeline required to bring a case to court. Therefore, the need for a system that can effectively and efficiently retrieve forensically relevant images has intensified.

Moreover, with each investigation being unique, the usability of pre-determined image categories is limited. It is therefore desirable to allow users to express their search intent and information needs freely and more thoroughly. Thus, this paper explores the potential of zero-shot cross-model retrieval, allowing retrieval across different modalities, such as querying images with natural language.

Due to the inherent limitations of retrieval systems and the difficulty of thoroughly understanding the user's search intent (e.g., due to an ambiguous query), it is nearly impossible to return satisfactory results for every query \cite{guo2018dialog, ma2014multiple}. A powerful way to improve retrieval performance is interactive learning \cite{xu2017relevance, rui1998relevance}, which was introduced to bring the user into the retrieval loop and bridge the semantic gap between what a query represents and what the user has in mind. In Excalibur, we apply interactive learning to improve image retrieval effectiveness in the forensic domain.

Finally, we experiment with different query strategies to explore the effect of the initial image retrieval on the effectiveness of interactive learning.

%% file: sections/statusquo.tex
\section{Current Situation and the Aim of the Proposed System}
The investigator's primary goal is to locate incriminating content. Increasingly, our devices serve as a looking glass into our lives. Therefore, investigations include exploring a suspect's seized device(s). Seizing a device is an intense invasion of privacy that is highly regulated by the prosecutor's office and is not done lightly.

As devices grow in storage capacity and overall usage, so does the amount of stored content. Generally, only a tiny portion of this content is relevant to the criminal investigation, which complicates achieving the aforementioned goal. Automated analysis of imagery strives to alleviate some of these impediments.

With Excalibur, we aim to provide an easy-to-use system for the investigators to comb through large amounts of images using intuitive querying quickly. Moreover, we strive for a system that works well in various fiscal- and criminal law domains since FIOD investigates various illegal activities, such as terrorist financing, money laundering, and drug trafficking, each requiring a different investigatory approach.

%% file: sections/related-work.tex
\section{Related work}
\label{related-work}

In recent years, a significant research effort was devoted to developing forensic tools to support investigators and relieve them from manual, time-consuming tasks \cite{gisolf2021search, bourouis2020recent, ibrahimi2021inside}. \cite{Park2015PartiallyOF} proposed a facial image retrieval system for partially occluded facial images. The proposed system was based on a non-statistical approach, avoiding the need for a supervised learning process. \cite{lee2011image} presented a content-based image retrieval system that can identify suspects based on their tattoos. The system automatically extracts features from a query image and retrieves near-duplicate tattoo images from an image database. While most approaches are rather specific, the goal of Excalibur is to increase the retrieval flexibility to allow the user to retrieve images of various visual concepts encountered in the forensic domain. 

Cross-modal retrieval has become an increasingly important research topic due to the rapid growth of unstructured multimedia data \cite{wang2016comprehensive}. Previously, image retrieval was restricted by a pre-determined set of semantic categories, allowing users to retrieve images using the category labels as queries \cite{smeulders2000content}. However, querying using natural language or an image enables a richer expression of the user's search intent. Cross-modal retrieval aims to take one type of modality as the query to retrieve semantically similar data from another type of modality.
To tackle the lack of flexibility in retrieving data from unseen classes, zero-shot cross-modal retrieval is an emerging research topic \cite{radford2021learning, yuan2021chop, ji2019attribute, long2018towards}.
Zero-shot approaches aim to perform when there is no training data available for some classes by leveraging semantic information within the data.

\noindent As mentioned before, querying with a pre-determined set of categories significantly limits the user to express their information need. Moreover, with each investigation being unique, it is infeasible to determine all possible domain-specific concepts and categories in advance. Therefore, in Excalibur, we leverage the power of zero-shot cross-modal retrieval to provide the user with a more flexible way of querying an image collection.

Interactive learning involves the iterative process of retrieving items from a collection and showing them to the user, who interacts with the system by rating the relevance of the retrieved items based on subjective judgments. The obtained relevance judgments are then used to re-rank the images, for instance, by training a classifier. Recently, interactive learning can perform at the scale of today's media collections \cite{khan2020interactive}. In addition, relevance feedback has proven to be a powerful mechanism for tasks that require alternating between search and exploration. Lastly, due to the advances in semantic data representations and high-dimensional indexing, interactive learning is receiving increased attention again \cite{khan2020interactive}.

%% file: sections/system-overview.tex
\begin{figure*}
\includegraphics[width=1.0\textwidth]{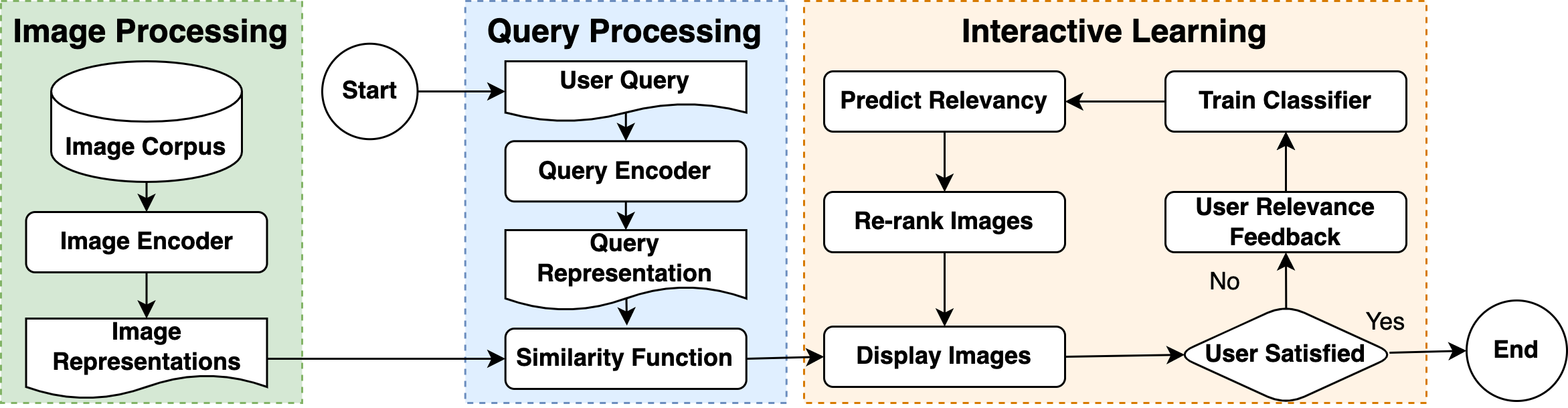}
\caption{A conceptual illustration of Excalibur.}
\label{fig:diagram}
\end{figure*}

\section{System Overview}
\label{system-overview}

In this section, we propose Excalibur - a zero-shot cross-modal image retrieval system extended with interactive learning. Figure \ref{fig:diagram} shows a conceptual overview of Excalibur. We follow the same notation as \cite{hendriksen2022extending}. Given a query $q$ and an image corpus $C$, the retrieval system returns a ranked list of images. The ranking is based on the output of the similarity function $S$. After the initial retrieval of the ranked list of images, we allow the user to refine the results by providing relevance feedback by marking images as relevant or non-relevant. The provided relevance feedback is used to train a binary classifier. Finally, the images are re-ranked based on the confidence of the predictions made by the trained classifier. 

\subsection{Image Processing}
\subsubsection{Image Encoder} The Image Encoder takes as input an image $i$ and returns its representation $h_i$. We use a pre-trained CLIP \cite{radford2021learning} model with a vision transformer-based backbone (ViT-B/32) to obtain the image representation vectors. CLIP is a neural network and jointly trains an image encoder and a text encoder to predict the correct image-text pairs, making it possible to compute the similarity between text and images. The choice for CLIP is motivated by the promising results that have been achieved on cross-modal retrieval \cite{zeng2022comprehensive, conde2021clip, hendriksen2022extending}. The most important aspect of CLIP is its zero-shot capabilities, allowing us to use the pre-trained model for our use case without any domain-specific fine-tuning.

\subsection{Query Processing}
\subsubsection{Query Strategy} 
We define two query strategies based on a user's domain expertise and knowledge of the image collection: \textit{Query with Natural Language} and \textit{Query with Image}. We elaborate on these strategies in the next section. 

\subsubsection{Query Encoder} The Query Encoder takes a query $q$ from a user, text, or an image and returns its representation $h_q$. We use the aforementioned image encoder for both the image $i$ and the query $q$. Since CLIP is trained on multiple modalities of data, it can be used to obtain representations for both modalities of a query. 

\subsubsection{Similarity Function}
The pre-trained CLIP model that Excalibur uses is trained to maximize the cosine similarity of the image and text embeddings of the correct pairs in the batch while minimizing the cosine similarity of the incorrect pairs \cite{radford2021learning}. Therefore, our similarity function $S$ also computes the cosine similarity between two representation vectors ($h_i, h_q$).

\subsection{Interactive Learning}
To improve the retrieval performance after the initial retrieval, we allow the user to provide relevance feedback about the retrieved images. The system subsequently uses this feedback to re-rank the images by relevance. The user can provide feedback in multiple interaction rounds until the information need is satisfied. More specifically, we follow the approach as outlined below:

\begin{enumerate}
    \setlength\itemsep{0.5em}
    \item The user selects $p$ relevant and $n$ non-relevant images from the search results.
    \item Train a binary classifier on the vector representations of the $p$ relevant and $n$ non-relevant images.
    \item Re-rank the images according to the predictions made by the trained classifier. The images are ranked based on the confidence of the predictions made by the trained classifier.
    \item Repeat steps (1) to (3) until the results are satisfactory.
\end{enumerate} 

Similar to other works in the field of user relevance feedback \cite{khan2020interactive, zahalka2017blackthorn}, the binary classifier used in Excalibur is a Support Vector Machine (SVM). The goal of the SVM is to find the margin that maximizes the distance between the representation vectors of the marked \textit{p} relevant and \textit{n} non-relevant images by the user. The choice for the SVM is motivated by the combination of the algorithm's performance, speed, and interpretability. Besides that, in contrast to deep learning algorithms, training an SVM requires relatively few labeled instances.

We implemented the binary classifier using SVC by Scikit-Learn \cite{scikit-learn}, and we experimented with the kernel and regularization parameter $C$. The other parameters remain at default.

\subsubsection{Human-in-the-loop}
Investigators are often looking for domain-specific concepts. By adding interactive learning to Excalibur, we provide a way for the user to interact with the system and give feedback that is incorporated in real-time. 

%% file: sections/experimental-setup.tex
\section{Experimental Setup}
\label{experimental-setup}

In this section, we first describe some of the core parameters and querying concepts. Then, we present our evaluation framework, including the dataset, protocol, and metrics. Lastly, we describe our user study.

\subsection{Experiments}

\subsubsection{System Parameters} 
We performed a grid search for the hyperparameters of the SVM classifier used in the interactive learning stage. Furthermore, for the simulation, we experiment with the number of positive and negative samples the user marks in each interaction round.

\paragraph{Query Strategy} 
As mentioned in the system overview, Excalibur supports image- and text-based querying. Often, investigators look for domain-specific concepts that are hard to express with natural language. Therefore, if an investigator has access to a similar image of interest, it is often more convenient to query the image collection using an image instead of describing the concept with natural language. We experiment with both these strategies to compare their performance. 

Since the pre-trained CLIP model is trained on 400 million image-text pairs found on the internet, we assume that querying in English works best. However, since Excalibur is aimed toward Dutch investigators, we also experiment with querying in Dutch. Furthermore, we experiment with adding a query prefix, as recommended in \cite{radford2021learning}, i.e., \textit{a photo of a \{concept\}} instead of solely \textit{\{concept\}}.
 
\subsection{Evaluation} The evaluation we apply to Excalibur is twofold. Firstly, we verify the retrieval performance of Excalibur by using a simulation protocol inspired by Analytic Quality \cite{zahalka2015analytic}. Secondly, we validate Excalibur via a user study with investigators on domain data.

\subsubsection{Simulation Setup} 

\noindent Frequently, investigators are tasked with ascertaining the temporal and geolocational components of digital evidence. Any evidence's time and location components are valuable and often serve as a burden of proof in court. Due to frequent absence of time and location metadata \cite{chen2019deep}, a higher-level representation, such as the scene (e.g., \textit{harbor}, \textit{parking lot}), is a more consistently available measure of the location. Moreover, such higher-level representations may serve as additional filters for the investigators. Therefore, for this simulated experiment, we consider the scenario where an investigator is challenged with the task of retrieving images related to a scene of interest.

\paragraph{Image Collection} Places365 \cite{zhou2017places} is one of the most exhaustive scene-centered datasets. We use the validation set of 100 images per scene category. The individual categories vary in scope - some are rather general (e.g., \textit{harbor}), while others are more specific (e.g., \textit{sushi bar}). The dataset for the experiment is constructed by selecting 20 out of the 365 scene categories in the Places365 dataset, where each scene is assigned to a simulated user; henceforth referred to as an \textit{actor}. The scene categories are selected in cooperation with investigators to ensure the scenes are forensically relevant (e.g., \textit{airport terminal} and \textit{hotel room}).

A caveat of using this dataset is that the semantics conveyed by the Places365 label for a particular image are not necessarily aligned with the user's interpretation. For instance, an image labeled as \textit{fast food restaurant} can be visually very similar to an image labeled as \textit{restaurant}, even though the labels of the images are distinct. Consequently, due to the distinct labels of the images, quantitative relevancy judgment contrasts with the user's relevancy judgment.  

\paragraph{Protocol}
After the initial retrieval, the actor selects $p$ positive and $n$ negative images to re-rank the search results. The actors base their relevancy judgment on the ground truth annotations provided by the Places365 dataset. The actors selected the positive and negative samples ranked highest in the list of images. 

To address the label caveat mentioned above, we compute the cosine similarity between the mean representation of the selected positive images and each possibly selected negative image. Determining the similarity threshold is an empirical question, which is not the focus of this work. If the cosine similarity is larger than $0.75$, we chose not to select the image as a negative sample.

Finally, similar to \cite{zahalka2020ii}, to simulate user behavior as close as possible and test the robustness of Excalibur, we implement an error rate in the relevancy judgments of the actors. For each image labeled by the actor, we sample a uniform random number $r \in [0,1]$. If $r > 0.8$, the actor makes one of the following mistakes with equal probability: ignores the image (i.e., provides no feedback at all) or flips relevance judgment (i.e., a non-relevant image will be marked as relevant, or vice versa).

\paragraph{Metric} At the end of an investigation, the evidence must be as complete as possible. Therefore, the first metric is $Recall$@$K$ where $K = 200$. We consider a relatively high value for $K$ since investigators are used to inspecting a large image collection during an investigation.

However, precision is often of greater importance during the early stages of an investigation than recall. Investigators require relevant information quickly to identify those devices that merit further investigation. Therefore, the second metric is Mean Average Precision@$K$ (\textit{MAP@K}), where $K = 50$. \textit{MAP@K} jointly considers the precision and ranking of images. We consider a relatively low value of $K$ since investigators lack time to go through multiple pages of results during the initial phase of an investigation. 

\subsubsection{User Study} \hfill

\noindent The user study aims to validate Excalibur on noisy - and possibly low-quality - images that require domain-specific knowledge. We aim to simulate the scenario where an investigator tries to determine whether there is incriminating content on a suspect's seized device. In total, five participating investigators completed the user study.

\begin{figure}
\includegraphics[width=0.45\textwidth]{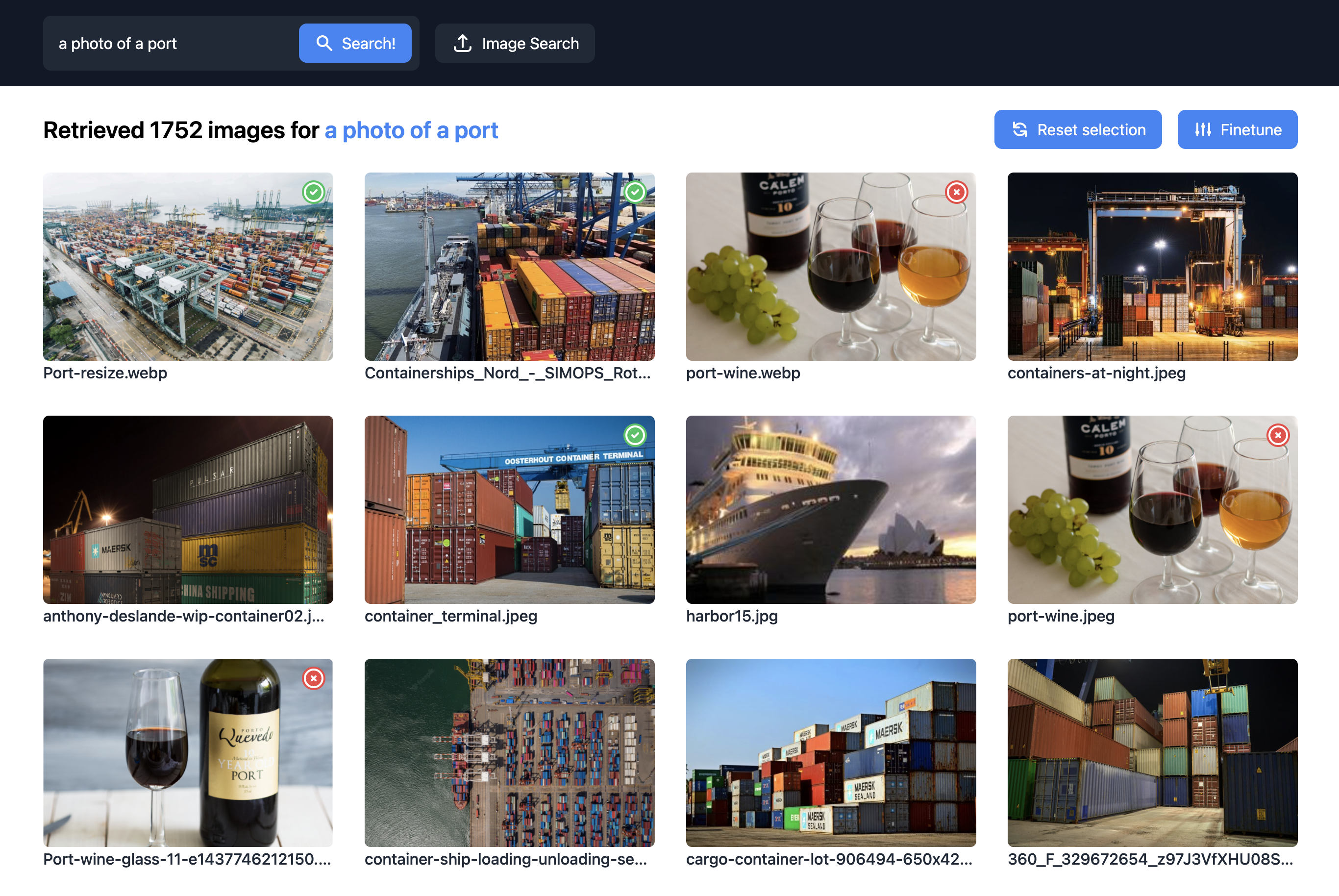}
\caption{Screenshot of the user interface of Excalibur.}
\label{fig:ui}
\end{figure}

\paragraph{User Interface} Figure \ref{fig:ui} shows the interface used during the study. The users could mark images as relevant or irrelevant for their search intent by clicking or shift-clicking the images, respectively. These choices are indicated using green and red badges in the top-right corner of the images. When the user is done selecting examples, they use the top-right \textit{Finetune} button to re-rank the search results. 

\paragraph{Image Collection} 
We worked with various datasets from multiple concluded investigations containing imagery representative of a typical investigation. Due to the sensitivity mentioned above, we are not allowed to share further details about the data.

\paragraph{Protocol} 

Firstly, we asked the participants questions concerning their current approach to searching and exploring through extensive image collections, after which we introduced Excalibur. Next, the participant is given 10 minutes to get familiar with Excalibur using their dataset.

Next, each participant performed ten unique search tasks, which were provided beforehand by the participants and were identical to the queries used during an investigation to prove a case in court. Finally, we asked the participant to complete the System Usability Scale \cite{brooke1996sus} questionnaire, consisting of ten questions concerning Excalibur's usability. In addition, we asked the participants to express their opinions on Excalibur's effectiveness, efficiency, and overall experience.

%% file: sections/results.tex
\section{Results}
\label{results}

\input{tables/main_results_alt_no_std}

\subsection{Simulation}

\subsubsection{System Parameters}
Table \ref{tab:system-params} shows the results of system parameter tuning. We used the most optimal configuration of the system for the simulations as well as the user study.

\input{tables/system_parameters.tex}

The negative multiplier denotes the number of selected negative samples in each interaction round. Similar to \cite{zahalka2020ii}, the number of negative samples is a multiple of the number of positive samples. Although we do not focus on the user experience, a lower value is preferred when the performance is similar.

A benefit of the non-linear kernels is the ability to capture multiple visual concepts simultaneously. On the other hand, non-linear kernels are slightly less efficient. However, we believe the benefits of a non-linear kernel outweigh the downsides.

The retrieval limit denotes the number of returned images during the initial retrieval. A higher initial retrieval limit increases the computational complexity of the interactive learning process. Therefore, a lower value for the initial retrieval limit is preferred when the performance is similar.

\begin{figure}[h]
  \centering
  \subfloat[a][\textit{MAP@50}]{\includegraphics[width=0.4\textwidth]{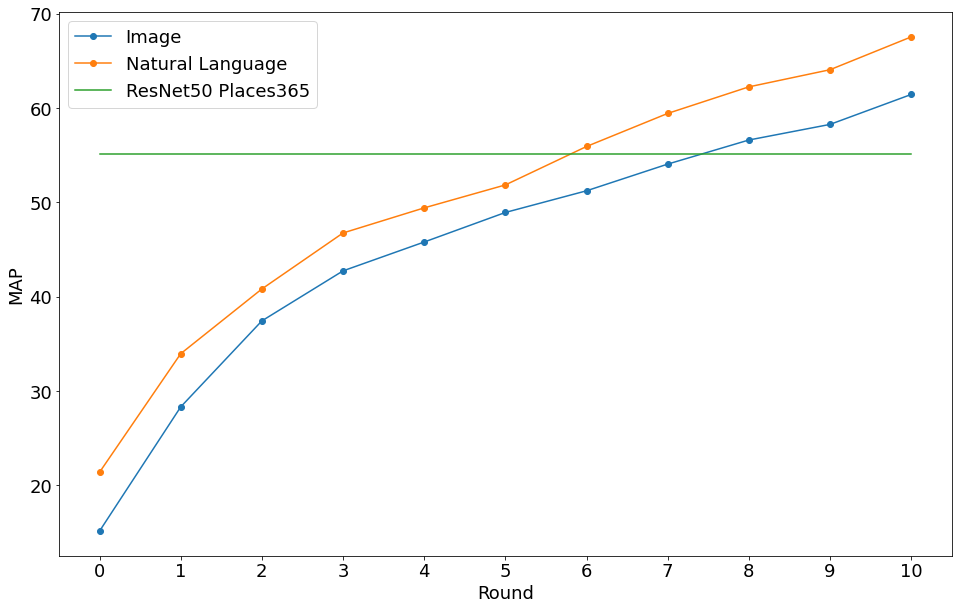} \label{fig:a}} \\
  \subfloat[b][$Recall$@$200$]{\includegraphics[width=0.4\textwidth]{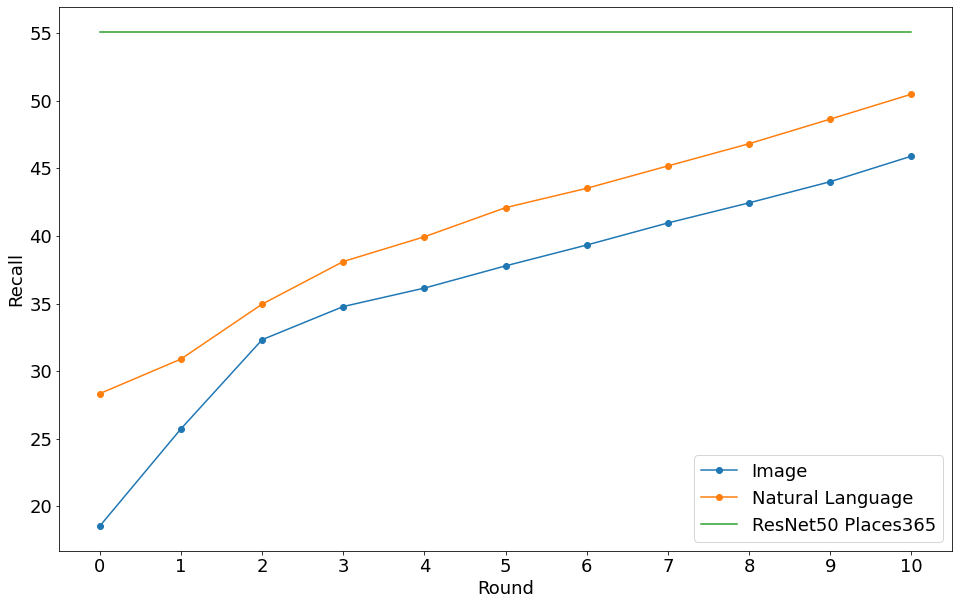} \label{fig:b}}
  \caption{Retrieval performance over ten interaction rounds for different query strategies.} \label{fig:main-results}
\end{figure}

\subsubsection{Query Strategy}

The results across ten interaction rounds using both query strategies are presented in Table \ref{tab:main-results}. To compare the strategies with a competitive supervised method, we used a pre-trained ResNet50 model that was trained on the Places356 dataset as a baseline. Figure \ref{fig:main-results} shows the \textit{MAP@50} and $Recall$@$200$ for different query strategies and the baseline model over ten interaction rounds. 

The zero-shot method with interactive learning approaches the performance of a tailor-made model in terms of \textit{Recall@200} and outperforms it in terms of \textit{MAP@50}. This indicates the system's ability to adapt to new domains, which is essential for the successful real-life adoption of Excalibur. We discuss the trade-off between a supervised model and our approach in more detail in the Discussion.

When looking at the performance of the initial retrieval, querying using natural language performs 42\% and 53\% better than querying with an image for \textit{MAP@50} and \textit{Recall@200}, respectively. Subsequently, we found that querying in English yields the most optimal retrieval performance. Additionally, using a prefix (i.e., adding \textit{``a photo of"} in front of the query) does not significantly improve the initial retrieval performance.

By leveraging user relevance feedback, after ten interaction rounds, the \textit{MAP@50} increased by 215\% when querying with natural language and 307\% when querying with an image. Moreover, the \textit{Recall@200} increased by 78\% and 148\% when querying with natural language and an image, respectively. We also found that querying using natural language results in more efficient interactive learning. That is, it requires fewer images to be viewed per interaction round to find a sufficient number of relevant samples.

\begin{figure}
\centering
\includegraphics[width=0.4\textwidth]{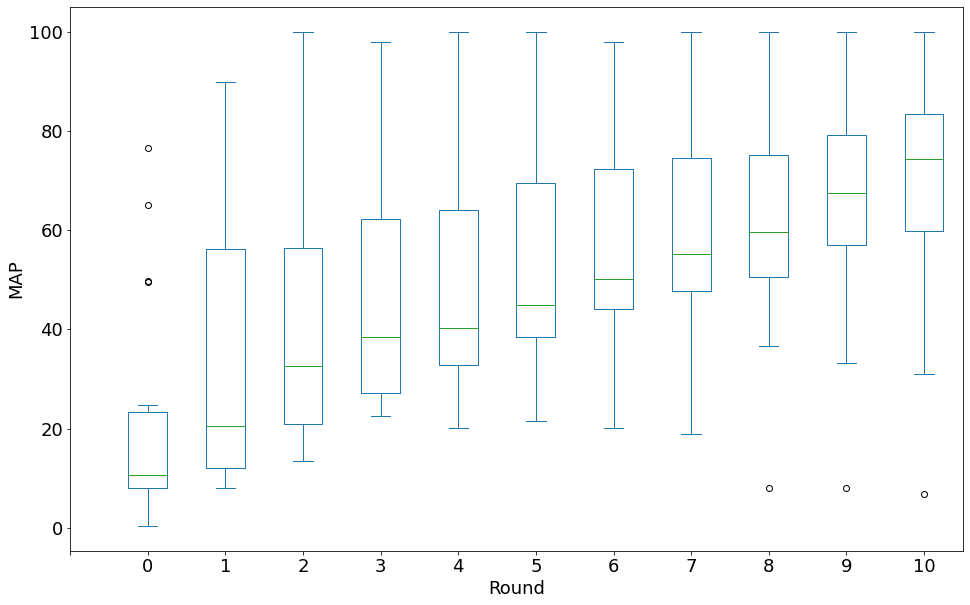}
\caption{MAP@50 of ten interactions with text query.}
\label{fig:boxplot}
\end{figure}

However, Figure \ref{fig:boxplot} shows a deviation in retrieval performance between the individual scenes while employing the natural language querying strategy. The deviations are similar when using the image querying strategy. We aim to understand the reason behind such differences in performance between different scenes in the Discussion.

\subsection{User Study}
The participants rated Excalibur 9.4/10.0 based on the System Usability Scale questionnaire. The participants preferred querying with an image when looking for a specific concept with domain-specific semantics (e.g., \textit{container seal}). 

Alternatively, when confronted with an unfamiliar image collection, they preferred querying with natural language to query with a broader concept or scene and afterward ``zoom in" by leveraging interactive learning. For example, the participants would first query with \textit{sea port} and then mark images of a \textit{container} as relevant. Overall, the participants considered Excalibur a substantial improvement over their current system and expressed interest in using it in their daily practice.

%% file: tables/main_results_alt_no_std.tex
% Please add the following required packages to your document preamble:
% \usepackage{multirow}
% \usepackage{graphicx}
\begin{table*}[]
\large
\centering
\resizebox{\textwidth}{!}{%
\begin{tabular}{|c|c|c|c|c|c|c|c|c|c|c|c|c|c|c|}
\hline
\textbf{Metric} &
  \textbf{Query Strategy} &
  \textbf{Initial retrieval} &
  \textbf{\% initial diff} &
  \textbf{Round 1} &
  \textbf{2} &
  \textbf{3} &
  \textbf{4} &
  \textbf{5} &
  \textbf{6} &
  \textbf{7} &
  \textbf{8} &
  \textbf{9} &
  \textbf{10} &
  \multicolumn{1}{c|}{\textbf{\% IL Gain}} \\ \hline
\textit{MAP@{50}} &
  Image &
  15.1 &
  \multirow{2}{*}{\begin{tabular}[c]{@{}c@{}}Natural Language \\ \textbf{42\%} \( \uparrow \) \end{tabular}} &
  28.3 &
  37.4 &
  42.7 &
  45.8 &
  48.9 &
  51.2 &
  54.0 &
  56.6 &
  58.3 &
  61.4 &
  \textbf{307\%}
   \\
 &
  Natural Language &
  21.4 &
   &
  33.9 &
  40.8 &
  46.7 &
  49.4 &
  51.8 &
  55.9 &
  59.4 &
  62.2 &
  64.1 &
  \textbf{67.5} &
  \textbf{215\%}
   \\ \hline
\textit{Recall@{200}} &
  Image &
  18.5 &
  \multirow{2}{*}{\begin{tabular}[c]{@{}c@{}}Natural Language \\ \textbf{53\%} \( \uparrow \) \end{tabular}} &
  25.7 &
  32.3 &
  34.8 &
  36.1 &
  37.8 &
  39.3 &
  41.0 &
  42.5 &
  44.0 &
  45.9 &
  \textbf{148\%}
   \\
 &
  Natural Language &
  28.3 &
   &
  30.9 &
  35.0 &
  38.1 &
  39.9 &
  42.1 &
  43.5 &
  45.2 &
  46.8 &
  48.7 &
  \textbf{50.5} &
  \textbf{78\%}
   \\ \hline
\end{tabular}%
}
\caption{Retrieval performance over 10 interaction rounds for different query strategies.}
  \label{tab:main-results}
\end{table*}

%% file: tables/system_parameters.tex
\begin{table}[H]
\small
\centering
\begin{tabular}{l|clll}
\toprule
\textbf{Parameter}                  & \multicolumn{1}{l}{\textbf{Optimal}} & \textbf{Grid}            \\ \hline
C                           & 10                                     & 0.1, 1, 10, 100         \\ 
Positive Samples        & 4                                     & 1, 2, 3, 4                 \\ 
Negative Multiplier & 2                                     & 1, 2, 3, 4                 \\ 
Kernel                     & RBF                                   & Linear, RBF \\
& & Poly, Sigmoid \\
Retrieval Limit & 2500                                   & 2500, 5000 \\ & & 7500, 10000                \\ 
\bottomrule
\end{tabular}
\caption{\label{tab:system-params}Results of the system parameter tuning.}
\end{table}

%% file: sections/path_to_deployment.tex
\section{Path to Deployment}
We will make a hosted version of Excalibur available to test at \url{https://excalibur-ui.herokuapp.com} shortly after the paper submission. In this demo version, we will allow the users to test the workings of the application using a demo dataset.

Precautions for security and privacy permeate the forensic domain, making an organization-wide deployment of the application a challenging task. Various deployment routes are available, each of which presents its risks and benefits. Given the sensitive nature of this data, we have opted for a local, on-premise deployment.

We see the need for Excalibur among investigators throughout the organization, and we plan to deploy the system across the regions where FIOD operates gradually. Moreover, we plan to host onboarding sessions with the investigators in each region to teach them how to use Excalibur in their investigations efficiently.

%% file: sections/discussion.tex
\section{Discussion}
\label{discussion}
There is a large deviation between the retrieval performance of the individual scenes. A possible explanation is that some scenes overlap with others, making them hard to recognize. For example, while the scene categories \textit{restaurant}, \textit{fast food restaurant}, and \textit{sushi bar} share many visual features, they fall under separate scene categories. Hence, when querying for photos of a \textit{restaurant}, photos with a ground truth of \textit{fast food restaurant} are considered wrong. We evaluated the performance on a relatively small subset of the Places365 scenes, and the results might not directly translate to the complete Places365 dataset. 

The pre-trained Places365 model is restricted to retrieving the pre-determined scene categories. The success of deep learning is heavily dependent on enormous amounts of labeled images, which are often very limited. Moreover, labeling images is a labor-intensive and expensive task. Thus, a supervised learning approach requires significant effort to adapt to a new task (i.e., predicting new visual concepts). Therefore, despite poorer initial retrieval performance, cross-modal learning offers a more flexible and scalable solution to retrieving images of various visual concepts encountered in the forensic domain and catches up to the supervised method using interactive learning. 

After the initial retrieval, there is a relatively large difference in retrieval performance between querying with natural language and images. However, after two interaction rounds, the gap in retrieval performance is reduced considerably. Therefore, we found that interactive learning is robust enough to work equally well with both query strategies. Interestingly, during the user study, we discovered that both query strategies serve their own purpose. 

%% file: sections/conclusion.tex
\section{Conclusion}
\label{conclusion}

We proposed Excalibur - a zero-shot cross-modal image retrieval system extended with interactive learning in this work. Excalibur enables search and exploration in large image collections by allowing users to refine the search results via relevance feedback. Excalibur can effectively retrieve images with the domain- and expertise-dependent semantics, making it suitable to be deployed in the forensic domain. The experimental results of the simulations confirm the effectiveness of interactive learning in improving retrieval performance after the initial retrieval. After ten interaction rounds, the \textit{MAP@50} and \textit{Recall@200} increased by 215\% and 78\%, respectively. Moreover, the user study participants were impressed by the retrieval performance of Excalibur and expressed interest in using it in their daily practice. Although we focused on improving retrieval performance in the forensic domain, many of our findings are easily transferable to other domains where searching vast amounts of images is required.

Based on this work, there are multiple directions future research could follow. Excalibur was developed to improve retrieval performance in the forensic domain. Meanwhile, scalability also becomes critical because investigations contain an increasing amount of unstructured data. To address that, Excalibur's computational efficiency and performance could be the focus of future work. For example, it could reduce the representation vectors' dimensionality.

Furthermore, future researchers could evaluate the simulation more robustly by adding relevancy-based metrics such as \textit{nDCG} or using a more consistently labeled dataset. Finally, future research could extend Excalibur by adding new functionalities. For example, investigators expressed their desire to filter on file size and extension and have metadata integrated into the search results.

%% file: main.bbl
\begin{thebibliography}{29}
\providecommand{\natexlab}[1]{#1}

\bibitem[{Battiato et~al.(2012)Battiato, Emmanuel, Ulges, and
  Worring}]{battiato2012multimedia}
Battiato, S.; Emmanuel, S.; Ulges, A.; and Worring, M. 2012.
\newblock Multimedia in forensics, security, and intelligence.
\newblock \emph{IEEE MultiMedia}, 19(1): 17--19.

\bibitem[{Bourouis et~al.(2020)Bourouis, Alroobaea, Alharbi, Andejany, and
  Rubaiee}]{bourouis2020recent}
Bourouis, S.; Alroobaea, R.; Alharbi, A.~M.; Andejany, M.; and Rubaiee, S.
  2020.
\newblock Recent advances in digital multimedia tampering detection for
  forensics analysis.
\newblock \emph{Symmetry}, 12(11): 1811.

\bibitem[{Brooke(1996)}]{brooke1996sus}
Brooke, J. 1996.
\newblock Sus: a “quick and dirty’usability.
\newblock \emph{Usability evaluation in industry}, 189(3).

\bibitem[{Chen and Davis(2019)}]{chen2019deep}
Chen, B.-C.; and Davis, L.~S. 2019.
\newblock Deep representation learning for metadata verification.
\newblock In \emph{2019 IEEE Winter Applications of Computer Vision Workshops
  (WACVW)}, 73--82. IEEE.

\bibitem[{Conde and Turgutlu(2021)}]{conde2021clip}
Conde, M.~V.; and Turgutlu, K. 2021.
\newblock CLIP-Art: contrastive pre-training for fine-grained art
  classification.
\newblock In \emph{Proceedings of the IEEE/CVF Conference on Computer Vision
  and Pattern Recognition}, 3956--3960.

\bibitem[{Gisolf, Geradts, and Worring(2021)}]{gisolf2021search}
Gisolf, F.; Geradts, Z.; and Worring, M. 2021.
\newblock Search and Explore Strategies for Interactive Analysis of Real-Life
  Image Collections with Unknown and Unique Categories.
\newblock In \emph{International Conference on Multimedia Modeling}, 244--255.
  Springer.

\bibitem[{Guo et~al.(2018)Guo, Wu, Cheng, Rennie, Tesauro, and
  Feris}]{guo2018dialog}
Guo, X.; Wu, H.; Cheng, Y.; Rennie, S.; Tesauro, G.; and Feris, R. 2018.
\newblock Dialog-based interactive image retrieval.
\newblock \emph{Advances in neural information processing systems}, 31.

\bibitem[{Hasham, Joshi, and Mikkelsen(2019)}]{hasham2019financial}
Hasham, S.; Joshi, S.; and Mikkelsen, D. 2019.
\newblock Financial crime and fraud in the age of cybersecurity.
\newblock \emph{McKinsey \& Company}.

\bibitem[{Hendriksen et~al.(2022)Hendriksen, Bleeker, Vakulenko, Noord, Kuiper,
  and Rijke}]{hendriksen2022extending}
Hendriksen, M.; Bleeker, M.; Vakulenko, S.; Noord, N.~v.; Kuiper, E.; and
  Rijke, M.~d. 2022.
\newblock Extending CLIP for Category-to-image Retrieval in E-commerce.
\newblock In \emph{European Conference on Information Retrieval}, 289--303.
  Springer.

\bibitem[{Ibrahimi et~al.(2021)Ibrahimi, van Noord, Alpherts, and
  Worring}]{ibrahimi2021inside}
Ibrahimi, S.; van Noord, N.; Alpherts, T.; and Worring, M. 2021.
\newblock Inside Out Visual Place Recognition.
\newblock \emph{arXiv preprint arXiv:2111.13546}.

\bibitem[{Ji et~al.(2019)Ji, Sun, Yu, Pang, and Han}]{ji2019attribute}
Ji, Z.; Sun, Y.; Yu, Y.; Pang, Y.; and Han, J. 2019.
\newblock Attribute-guided network for cross-modal zero-shot hashing.
\newblock \emph{IEEE transactions on neural networks and learning systems},
  31(1): 321--330.

\bibitem[{Khan et~al.(2020)Khan, J{\'o}nsson, Rudinac, Zah{\'a}lka,
  Ragnarsd{\'o}ttir, {\TH}orleiksd{\'o}ttir, Gu{\dh}mundsson, Amsaleg, and
  Worring}]{khan2020interactive}
Khan, O.~S.; J{\'o}nsson, B.~{\TH}.; Rudinac, S.; Zah{\'a}lka, J.;
  Ragnarsd{\'o}ttir, H.; {\TH}orleiksd{\'o}ttir, {\TH}.; Gu{\dh}mundsson,
  G.~{\TH}.; Amsaleg, L.; and Worring, M. 2020.
\newblock Interactive learning for multimedia at large.
\newblock In \emph{European Conference on Information Retrieval}, 495--510.
  Springer.

\bibitem[{Lee et~al.(2011)Lee, Jain, Tong et~al.}]{lee2011image}
Lee, J.; Jain, A.; Tong, W.; et~al. 2011.
\newblock Image retrieval in forensics: tattoo image database application.
\newblock \emph{IEEE MultiMedia}, 19(1): 40--49.

\bibitem[{Long et~al.(2018)Long, Liu, Shen, and Shao}]{long2018towards}
Long, Y.; Liu, L.; Shen, Y.; and Shao, L. 2018.
\newblock Towards affordable semantic searching: Zero-shot retrieval via
  dominant attributes.
\newblock In \emph{Proceedings of the AAAI Conference on Artificial
  Intelligence}, volume~32.

\bibitem[{Ma and Lin(2014)}]{ma2014multiple}
Ma, Y.; and Lin, H. 2014.
\newblock A multiple relevance feedback strategy with positive and negative
  models.
\newblock \emph{PloS one}, 9(8): e104707.

\bibitem[{Park et~al.(2015)Park, Lee, Yoo, Kim, and Kim}]{Park2015PartiallyOF}
Park, S.; Lee, H.; Yoo, J.-H.; Kim, G.; and Kim, S.-N. 2015.
\newblock Partially Occluded Facial Image Retrieval Based on a Similarity
  Measurement.
\newblock \emph{Mathematical Problems in Engineering}, 2015: 1--11.

\bibitem[{Pedregosa et~al.(2011)Pedregosa, Varoquaux, Gramfort, Michel,
  Thirion, Grisel, Blondel, Prettenhofer, Weiss, Dubourg, Vanderplas, Passos,
  Cournapeau, Brucher, Perrot, and Duchesnay}]{scikit-learn}
Pedregosa, F.; Varoquaux, G.; Gramfort, A.; Michel, V.; Thirion, B.; Grisel,
  O.; Blondel, M.; Prettenhofer, P.; Weiss, R.; Dubourg, V.; Vanderplas, J.;
  Passos, A.; Cournapeau, D.; Brucher, M.; Perrot, M.; and Duchesnay, E. 2011.
\newblock Scikit-learn: Machine Learning in {P}ython.
\newblock \emph{Journal of Machine Learning Research}, 12: 2825--2830.

\bibitem[{Radford et~al.(2021)Radford, Kim, Hallacy, Ramesh, Goh, Agarwal,
  Sastry, Askell, Mishkin, Clark et~al.}]{radford2021learning}
Radford, A.; Kim, J.~W.; Hallacy, C.; Ramesh, A.; Goh, G.; Agarwal, S.; Sastry,
  G.; Askell, A.; Mishkin, P.; Clark, J.; et~al. 2021.
\newblock Learning transferable visual models from natural language
  supervision.
\newblock In \emph{International Conference on Machine Learning}, 8748--8763.
  PMLR.

\bibitem[{Rui et~al.(1998)Rui, Huang, Ortega, and Mehrotra}]{rui1998relevance}
Rui, Y.; Huang, T.~S.; Ortega, M.; and Mehrotra, S. 1998.
\newblock Relevance feedback: A power tool for interactive content-based image
  retrieval.
\newblock \emph{IEEE Transactions on circuits and systems for video
  technology}, 8(5): 644--655.

\bibitem[{Smeulders et~al.(2000)Smeulders, Worring, Santini, Gupta, and
  Jain}]{smeulders2000content}
Smeulders, A.~W.; Worring, M.; Santini, S.; Gupta, A.; and Jain, R. 2000.
\newblock Content-based image retrieval at the end of the early years.
\newblock \emph{IEEE Transactions on pattern analysis and machine
  intelligence}, 22(12): 1349--1380.

\bibitem[{Wang et~al.(2016)Wang, Yin, Wang, Wu, and
  Wang}]{wang2016comprehensive}
Wang, K.; Yin, Q.; Wang, W.; Wu, S.; and Wang, L. 2016.
\newblock A comprehensive survey on cross-modal retrieval.
\newblock \emph{arXiv preprint arXiv:1607.06215}.

\bibitem[{Worring, Engl, and Smeria(2012)}]{worring2012multimedia}
Worring, M.; Engl, A.; and Smeria, C. 2012.
\newblock A multimedia analytics framework for browsing image collections in
  digital forensics.
\newblock In \emph{Proceedings of the 20th ACM international conference on
  Multimedia}, 289--298.

\bibitem[{Xu, Wang, and Mao(2017)}]{xu2017relevance}
Xu, H.; Wang, J.-y.; and Mao, L. 2017.
\newblock Relevance feedback for Content-based Image Retrieval using deep
  learning.
\newblock In \emph{2017 2nd International Conference on Image, Vision and
  Computing (ICIVC)}, 629--633. IEEE.

\bibitem[{Yuan et~al.(2021)Yuan, Wang, Chen, and Zhong}]{yuan2021chop}
Yuan, X.; Wang, G.; Chen, Z.; and Zhong, F. 2021.
\newblock CHOP: An orthogonal hashing method for zero-shot cross-modal
  retrieval.
\newblock \emph{Pattern Recognition Letters}, 145: 247--253.

\bibitem[{Zah{\'a}lka et~al.(2017)Zah{\'a}lka, Rudinac, J{\'o}nsson, Koelma,
  and Worring}]{zahalka2017blackthorn}
Zah{\'a}lka, J.; Rudinac, S.; J{\'o}nsson, B.~{\TH}.; Koelma, D.~C.; and
  Worring, M. 2017.
\newblock Blackthorn: large-scale interactive multimodal learning.
\newblock \emph{IEEE Transactions on Multimedia}, 20(3): 687--698.

\bibitem[{Zah{\'a}lka, Rudinac, and Worring(2015)}]{zahalka2015analytic}
Zah{\'a}lka, J.; Rudinac, S.; and Worring, M. 2015.
\newblock Analytic quality: evaluation of performance and insight in multimedia
  collection analysis.
\newblock In \emph{Proceedings of the 23rd ACM international conference on
  Multimedia}, 231--240.

\bibitem[{Zah{\'a}lka, Worring, and Van~Wijk(2020)}]{zahalka2020ii}
Zah{\'a}lka, J.; Worring, M.; and Van~Wijk, J.~J. 2020.
\newblock II-20: Intelligent and pragmatic analytic categorization of image
  collections.
\newblock \emph{IEEE Transactions on Visualization and Computer Graphics},
  27(2): 422--431.

\bibitem[{Zeng and Mao(2022)}]{zeng2022comprehensive}
Zeng, Z.; and Mao, W. 2022.
\newblock A Comprehensive Empirical Study of Vision-Language Pre-trained Model
  for Supervised Cross-Modal Retrieval.
\newblock \emph{arXiv preprint arXiv:2201.02772}.

\bibitem[{Zhou et~al.(2017)Zhou, Lapedriza, Khosla, Oliva, and
  Torralba}]{zhou2017places}
Zhou, B.; Lapedriza, A.; Khosla, A.; Oliva, A.; and Torralba, A. 2017.
\newblock Places: A 10 million image database for scene recognition.
\newblock \emph{IEEE transactions on pattern analysis and machine
  intelligence}, 40(6): 1452--1464.

\end{thebibliography}
